\documentclass[preprint,showpacs,preprintnumbers,amsmath,amssymb,nofootinbib]{revtex4}

\usepackage{graphicx}
\usepackage{dcolumn}
\usepackage{bm}

\newcommand{\bi}{\bibitem}
\newcommand{\nn}{\nonumber}

\newcommand{\be}{\begin{eqnarray}}
\newcommand{\ee}{\end{eqnarray}}
\catcode`\@=11
\def\lsim{\mathrel{\mathpalette\@versim<}}
\def\gsim{\mathrel{\mathpalette\@versim>}}
\def\@versim#1#2{\vcenter{\offinterlineskip
\ialign{$\m@th#1\hfil##\hfil$\crcr#2\crcr\sim\crcr } }}
\catcode`\@=12

\begin{document}

\preprint{MPP-2004-20, KANAZAWA-04-05}

\title{Higgs Potential  in Minimal $S_3$ 
Invariant Extension \\ of the Standard Model}

\author{Jisuke Kubo$^{1,2}$}
\author{Hiroshi Okada$^{2}$}
\author{Fumiaki  Sakamaki$^{2}$}

\affiliation{
$^{1}$ Max-Planck-Institut f\"ur Physik,
 Werner-Heisenberg-Institut,
D-80805 Munich, Germany\\
$^2$Institute for Theoretical Physics, Kanazawa
University, Kanazawa 920-1192, Japan
 }

\vspace{2cm}

\begin{abstract}
 Minimal $S_{3}$ invariant Higgs potential with real
 soft $S_{3}$ breaking
 masses is investigated.
 It is required  that
without having a problem
with  triviality,
 all  physical Higgs bosons,
 except one neutral one,
become heavy  $\gsim 10$ TeV in order to
sufficiently suppress  flavor changing neutral currents.
There exist three nonequivalent soft mass terms that can be 
characterized according
to their discrete symmetries, and the one  which breaks
$S_{3}$ completely.
The $S_{2}'$ invariant
vacuum expectation values (VEVs)  of the Higgs fields are the most economic VEVs
in the sense that
the freedom of VEVs 
can be completely absorbed into the Yukawa couplings
so that it is possible to derive, without referring to the details 
of the VEVs,  the most general form for the fermion mass matrices
in  minimal $S_{3}$ extension of the standard model.
We find that except for the completely broken  case of the soft terms,
 the $S_{2}'$ invariant VEVs are unique
VEVs that satisfy the requirement of heavy Higgs bosons.
It is  found that they also correspond to a local minimum in
the completely broken  case.
 
\end{abstract}
\pacs{11.30.Hv, 12.60.Fr,12.15.Ff}

\maketitle

\section{Introduction}
A nonabelian flavor symmetry is certainly a powerful tool
to understand  flavor physics.
In the  case of the standard model (SM), where only
one  Higgs $SU(2)_{L}$ doublet  is present,
any nonabelian flavor symmetry has to be explicitly broken to
describe experimental data.
However, if the Higgs sector is extended, and 
 Higgs fields belong to
a nontrivial representation of a flavor group \cite{pakvasa1,pakvasa2},
phenomenologically viable possibilities may arise.
The  smallest nonabelian
discrete group is  $S_{3}$
\footnote{Flavor symmetries based on a 
permutation symmetry have been considered 
by many authors in the past. One of the first papers
on permutation symmetries  are
\cite{pakvasa1,pakvasa2,yamanaka,harari,koide1}.
See \cite{fritzsch} for a review.
Phenomenologically
viable models based on 
nonabelian discrete flavor symmetries $S_{3}, D_{4}$ and $A_{4}$ 
and also on a product of abelian discrete symmetries 
 have been recently constructed in
 \cite{kubo1,kubo2,kobayashi1,choi,ma5},  \cite{grimus2},  
 \cite{ma1,babu1,babu2,hirsch} and \cite{grimus1,ma4}, respectively.
(See also  \cite{koide3,koide2,ohlsson,kitabayashi,grimus3}. )
However, it is difficult 
to understand bi-large mixing of neutrinos
in terms of abelian discrete symmetries alone \cite{low1}.}.
It is a permutation group of three objects, and
offers a possible explanation why there are three
generations of the quarks and leptons \cite{kubo1,kubo2}.
An $S_{3}$ invariant Yukawa sector of the SM
has  exactly five independent couplings \cite{kubo1,kubo2}
\be
1 &:& L_{a}R_{a}H_{a}+L_{b}R_{b}H_{b}+L_{c}R_{c}H_{c},\nn\\
2 &:& L_{a}(R_{b}+R_{c})H_{a}+L_{b}(R_{a}+R_{c})H_{b}
+L_{c}(R_{a}+R_{b})H_{c},\nn\\
3 &:&( L_{b}+L_{c}) R_{a}H_{a}+( L_{a}+L_{c} )R_{b}H_{b}
+( L_{a}+L_{b})R_{c}H_{c},\\
4 &:&( L_{b}R_{b}+L_{c}R_{c})H_{a}+( L_{a}R_{a}+L_{c}R_{c})H_{b}+
( L_{a}R_{a}+L_{b}R_{b})H_{c},\nn\\
5 &:&( L_{b}R_{c}+L_{c}R_{b})H_{a}+( L_{a}R_{c}+L_{c}R_{a})H_{b}+
( L_{a}R_{b}+L_{b}R_{a})H_{c},\nn
\ee
where $L_{a}$, $R_{a}$ and $H_{a}$ correspond to three
left-handed leptons, right-handed leptons and  Higgs bosons,
which are subject to permutations. The three dimensional
representation ${\bf 3}$ of $S_{3}$ is not an irreducible representation;
${\bf 3}$ can be decomposed into ${\bf 1}$ and ${\bf 2}$ as
\be
{\bf 1} &:& H_S=\frac{1}{\sqrt{3}}(H_a+H_b+H_c),\\
{\bf 2} &:&(H_{1},H_{2})=(~\frac{1}{\sqrt{2}}(H_a-H_b),
\frac{1}{\sqrt{6}}(H_a+H_b-2H_c)~),
\ee
and similarly for $L$'s and $R$'s. In terms of the fields in the irreducible basis, the 
five independent Yukawa couplings are  \cite{kubo1,kubo2}:
\be
L_i R_i H_S,& &f_{ijk}L_i R_j H_k,~ L_S R_S H_S,~ L_S R_i H_i,~ L_i R_S H_i,
\label{yukawa}
\ee
where $i,j,k$ run from $1$ to $2$, and 
\be
f_{112}=f_{121}=f_{211}=-f_{222}=1.
\label{fijk}
\ee
It has been found in  \cite{kubo1,kubo2} that these Yukawa couplings are sufficient 
to reproduce the masses of the quarks and their mixing,
and that they are not only consistent with the known
observations in the leptonic sector, but also  can make 
testable predictions in the neutrino sector
if one assumes an additional discrete symmetry
in this sector.
In deriving the fermion mass matrices, it has been assumed  in  \cite{kubo1,kubo2} that
the vacuum expectation values (VEVs) 
of the Higgs fields are $S_{2}'$ invariant, i.e.
\be
<H_{S}> &\neq &0, ~<H_{1}>=<H_{2}>\neq 0.
\label{s2vev}
\ee
By the $S_{2}'$ invariance we mean an invariance
under the interchange of $H_{1}$ and $H_{2}$, i.e.
\be
H_{1} &\leftrightarrow & H_{2}.
\label{s2p}
\ee
Note that this permutation symmetry is not a subgroup of the original $S_{3}$.
Although the Yukawa couplings (\ref{yukawa}) do not respect this symmetry,
each term in  the $S_{3}$ invariant Higgs potential (given in (\ref{vH}))
except for one term respects this discrete symmetry.
Moreover, as we can see from (\ref{yukawa}), the $S_{2}'$ invariant
 VEVs (\ref{s2vev}) are the most economic VEVs
in the sense that
the freedom of VEVs 
can be completely absorbed into the Yukawa couplings
so that we can derive the most general form for the fermion mass matrices
\be
{\bf M} = \left( \begin{array}{ccc}
m_1+m_{2} & m_{2} & m_{5} 
\\  m_{2} & m_1-m_{2} &m_{5}
  \\ m_{4} & m_{4}&  m_3
\end{array}\right)
\label{general-m}
\ee
without referring to the details of VEVs.
In other words, if $<H_{1}>\neq <H_{2}>$, the mass matrices would have
one more independent parameter which should be determined in the Higgs
sector.

In the present paper we would like to investigate
how unique the $S_{2}'$ invariant vacuum is under the requirement that
except one neutral physical Higgs boson all the physical Higgs bosons can
become heavy $\gsim 10$ TeV without having a problem
with triviality \cite{luescher}.
This  bound  results in order to suppress 
three-level flavor changing neutral currents (FCNCs) 
that contribute, for instance, to the mass difference 
$\Delta m_{K}$ of $K^{0}$ and $\overline{K}^{0}$
 in $S_{3}$ invariant extension of the SM \cite{kubo3,okada}.
 [See also Ref. \cite{yamanaka} and \cite{brown}.]
The investigations are presented  in Sect. 3 and 4, and 
the conclusions are summarized  in the last section.
In Sect. V we discuss the Pakvasa-Sugawara vacuum \cite{pakvasa1}, and 
a supersymmetric case is treated in Sect. 6.

\section{$S_{3}$ invariant Higgs potential and soft $S_{3}$ breaking}
\subsection{$S_{3}$ invariant Higgs potential and its problem}
The  most general, $S_{3}$ invariant, renormalizable  potential
is given by \cite{pakvasa1}
\be
V_H &=&V_{2H}+V_{4H},\label{vH}\\
V_{2H} &=&-\mu_1^2(H_1^{\dagger}H_1+H_2^{\dagger}H_2)-
\mu_3^2 H_S^{\dagger}H_S, \label{2H}\nn\\
V_{4H} &= &+\lambda_{1}(H_1^{\dagger}H_1+H_2^{\dagger}H_2)^2
+\lambda_{2}(H_1^{\dagger}H_2-H_2^{\dagger}H_1)^2 \nn\\
& &+\lambda_{3}((H_1^{\dagger}H_2+H_2^{\dagger}H_1)^2
+(H_1^{\dagger}H_1-H_2^{\dagger}H_2)^2)\nn\\
& &+\{\lambda_{4}f_{ijk}(H_S^{\dagger}H_i)(H_j^{\dagger}H_k)+h.c.\}
+\lambda_{5}(H_S^{\dagger}H_S)
(H_1^{\dagger}H_1+H_2^{\dagger}H_2)\nn\\
& &+\lambda_{6}\{(H_S^{\dagger}H_1)(H_1^{\dagger}H_S)
+(H_S^{\dagger}H_2)(H_2^{\dagger}H_S)\}\nn\\
& &+\{\lambda_{7}[(H_S^{\dagger}H_1)(H_S^{\dagger}H_1)
+(H_S^{\dagger}H_2)(H_S^{\dagger}H_2)]+h.c.\}\nn\\
 & &+\lambda_{8}(H_S^{\dagger}H_S)^2,
 \label{v4H}
 \ee
 where $\lambda_{4}$ and $\lambda_{7}$ 
 can be complex \footnote{The $S_{3}$ invariant potential 
has been  studied in \cite{pakvasa1,koide3}, for instance.
Similar potentials with nonabelian
discrete symmetries have been 
 also studied in \cite{pakvasa2,yamanaka, grimus2, babu3}.}.
We first redefine $H_{i}$ as
\be
H_{\pm} &=&\frac{1}{\sqrt{2}}(H_1\pm H_2),
\ee
and write the $SU(2)_{L}$ Higgs doublets in components:
\be
H_{\pm} &=& 
 \left( \begin{array}{c}
h_{\pm} +i \chi_{\pm}\\
\frac{1}{\sqrt{2}} (h_{\pm}^{0}
  +i \chi_{\pm}^{0}) \\
\end{array}\right),~
H_S =
 \left( \begin{array}{c}
h_S +i \chi_S\\
 \frac{1}{\sqrt{2}} (h_S^0+i \chi_S^0) \\
\end{array}\right).
\ee
The down components of the Higgs doublets have zero electric charge,
and therefore, we assume that only the down components can
acquire a VEV.
Further, because of $U(1)_{Y}$ gauge invariance, 
it is always possible to make a phase rotation for  $H_{S}$ 
so that only the  real part  $h_S^0$ can get VEV.
We denote the VEVs as follows:
\be
<h_{\pm}^{0}> &=&v_{\pm},~ <h_S^0>=v_{S}~,
<\chi_{\pm}>=c_{\pm}~,
\label{vevs}
\ee
which should satisfy the constraint
\be
(v_{+}^{2}+v_{-}^{2}+v_{S}^{2}+c_{+}^{2}+c_{-}^{2})^{1/2}
&=& v \simeq 246 ~~\mbox{GeV}.
\label{constraint1}
\ee
In order to reproduce realistic fermion masses and  their mixings \cite{kubo1},
we also require that
\be
v_{S} &\neq& 0,~~\mbox{and at least one of}
~v_{\pm}~\mbox{and}~c_{\pm}~\neq 0
\label{require1}
\ee
is satisfied, and do not allow a large hierarchy among the 
nonvanishing VEVs,
unless it is noticed.
[In sections II and III, we however allow such hierarchy.]

There are five minimization conditions:
\be
0 &=& -v_{S}\mu_{3}^{2}+\partial V_{4H}/\partial h_{S}^{0},
\label{m30}\\
0 &=& -v_{+}\mu_{1}^{2}+\partial V_{4H}/\partial h_{+}^{0},
\label{vp0}\\
0 &=&- v_{-}\mu_{1}^{2}+\partial V_{4H}/\partial h_{-}^{0},
\label{vm0}\\
0 &=& -c_{+}\mu_{1}^{2}+\partial V_{4H}/\partial \chi_{+}^{0},
\label{cp0}\\
0 &=& -c_{-}\mu_{1}^{2}+\partial V_{4H}/\partial \chi_{-}^{0},
\label{cm0}
\ee
We regard VEVs as independent parameters, and  express
the parameters of the potential (\ref{vH}),
especially 
the mass parameters $\mu_{1}^{2}$ and $\mu_{3}^{2}$, 
in terms of
the VEVs. To make all the physical Higgs bosons except one neutral 
Higgs boson
without having large values of the Higgs quartic couplings $\lambda$'s,
we have to have either $-\mu_3^2, -\mu_1^2 >> v^2$ or
 $ -\mu_1^2 >> v^2$, where $v$ is defined in (\ref{constraint1}).
  For the first case, none of the VEVs can be
 $O(v)$,  because the derivative terms,
 i.e. $\partial V_{4H}/\partial h_{+}^{0}$ etc,
 are of $O(\mbox{VEV}^3)$.
 Therefore,
this case can not satisfy the constraint (\ref{constraint1}).
 For the second case,  $\mu_3$ and $v_S$ can be $O(v)$, but none of
 $v_+,v_-,c_+,c_-$ can be $O(v)$. That is, 
 the hierarchy $|v_+/v_S|, |v_-/v_S|, |c_+/v_S|, |c_-/v_S| << 1$
 has to be satisfied. This hierarchy is consistent with the minimization
 conditions (\ref{vp0})--(\ref{cm0}), only  if 
 at least one of the derivative terms,
 i.e. $\partial V_{4H}/\partial h_{+}^{0}$ etc, contains at least
 a term proportional to 
 $v_S^3$.  However, this is not the case, as we can see from the potential
 $V_{4H}$ (\ref{v4H}).
 Moreover, (\ref{require1}) does not allow $v_+=v_-=c_+=c_-=0$.

 It is thus clear, if the two conditions (\ref{constraint1}) and 
 (\ref{require1}) are satisfied,  that $\mu_{1}^{2}, \mu_{3}^{2}
\sim O(\mbox{VEV}^{2})$, which means that 
all the masses of the physical Higgs bosons are of $O(\mbox{VEV})$.
That is, to have a large Higgs mass, the
value of certain Higgs couplings $\lambda$'s have to be large.
Then we are running into the problem with triviality;
the Higgs mass can not be larger than the cutoff.
As we see from (\ref{vH}), the model has 
many Higgs couplings, so  that the known triviality
bound on the Higgs mass,
$\sim 700$ GeV \cite{luescher}, can not be directly applied .
But we may assume that the bound for the present case does not
differ very much from  that of the SM.
However, this upper bound  is too law  to suppress 
three-level flavor changing neutral currents (FCNCs) 
that contribute, for instance, to the mass difference 
$\Delta m_{K}$ of $K^{0}$ and $\overline{K}^{0}$;
certain Higgs masses in $S_{3}$ invariant extension of the SM
have to be larger than $\sim O(10)$ TeV \cite{yamanaka,kubo3,okada}.
Therefore,   in a phenomenologically
viable $S_{3}$ extension
of the SM,  $S_{3}$ symmetry
should be broken,
unless there is some cancellation mechanism
of FCNCs.

\subsection{Soft $S_{3}$ breaking  terms and their characterization}
As we have seen above, we have to modify 
the Higgs potential (\ref{vH})  to make it possible that
the Higgs masses can become lager than $10$ TeV.
How should we break $S_{3}$?
We would like to maintain the consistency and
predictions of $S_{3}$ in the Yukawa sector, while
 satisfying simultaneously the experimental constraints from 
FCNC phenomena.
Therefore, we  break $S_{3}$ as soft as possible.
The softest operators in the case at hand are those
of dimension  two;  that is, mass terms.
There are four soft breaking mass terms
\be
V_{SB}& =&-\mu_{2}^2(H_{+}^{\dagger}H_{+}-H_{-}^{\dagger}H_{-})
-\sqrt{2}(\mu_{4}^2H_{S}^{\dag}H_{+}+h.c.)\nn\\
& &-(\mu_{5}^{2}H_{+}^{\dagger}H_{-}+h.c.)
-\sqrt{2}(\mu_{6}^{2} H_{S}^{\dag}H_{-}+h.c.).
\label{vSB}
\ee
$\mu_{4}^{2}, \mu_{5}^{2}$ and $ \mu_{6}^{2}$ can be complex
parameters \footnote{The soft mass terms (\ref{vSB})
may be generated from a $S_3$ invariant Higgs potential
by introducing certain $S_3$ singlet Higgs fields \cite{brown}.}.
However,  we assume that
they are real parameters in following discussions except in Sect. 5.
We would like to  characterize these four mass terms according to 
discrete symmetries:
\be
R &:& H_{S} \to -H_{S}, 
\label{discrete1}\\
S_{2}' &:& H_{-} \to  -H_{-},
\label{discrete2}\\
S_{2}'' &:& H_{+} \to  -H_{+},\\
R\times S_{2}' &:& H_{S} \to -H_{S}~\mbox{and}~ H_{-} \to  -H_{-},
\label{discrete3}\\
R\times S_{2}''&:& H_{S} \to -H_{S} ~\mbox{and}~H_{+} \to  -H_{+}\\
S_{2}'\times S_{2}'' &:& H_{-} \to  -H_{-} ~\mbox{and}~ H_{+} \to  -H_{+},
\ee
where $S_{2}'$ and $S_{2}''$ are not a subgroup of the original $S_{3}$.
Accordingly, 
we characterize the soft mass terms (\ref{vSB})  as
\be
R &:&  \mu_{4}=\mu_{6}=0,
\label{class1}\\
S_{2}' &:& \mu_{5}=\mu_{6}=0,\label{class2}\\
S_{2}'' &:&\mu_{4}=\mu_{5}=0, \label{class4}\\
R\times S_{2}' &:& \mu_{4}=\mu_{5}=\mu_{6}=0,\label{class3}\\
R\times S_{2}''&:& \mu_{4}=\mu_{5}=\mu_{6}=0,\\
S_{2}'\times S_{2}'' &:& \mu_{4}=\mu_{5}=\mu_{6}=0.
 \label{classes}
\ee
Actually, there are only four nonequivalent soft breaking mass terms,
including one without no discrete symmetry.
This is because $S_{2}'$  and $S_{2}''$ are not independent:
The Higgs potential (\ref{vH}) and the soft terms (\ref{vSB})
are invariant under the interchange of  $H_{+}$ and $H_{-}$
if one appropriately redefines the coupling constants and mass parameters.
In the next section we will discuss the three cases, i.e.
$R,  S_{2}' $ and $R\times S_{2}'$ invariant cases, and in Sect.~4
we will treat the completely broken case, in which  all the soft 
mass terms (\ref{vSB}) are present.
Each possibility is renormalizable, because all the other interactions 
are $S_{3}$ invariant and can 
not induce infinite $S_{3}$ violating breaking terms (\ref{vSB}).
In principle, $\mu_{4}^{2},\mu_{5}^{2}$ and $\mu_{6}$ can be complex.
As announced, however, we assume that they are real, except for Sect. 5.
This is consistent with renormalizability from the same reason above.

Before we go to the
next sections, it may be worthwhile to write down explicitly 
the $\lambda_{4}$ and $\lambda_{7}$
terms of the potential $V_{4H}$ (\ref{v4H}):
\be
2\sqrt{2}V_{\lambda_{4}H}
&=&(\mbox{Re}(\lambda_{4})\chi_{S}^{0}-
\mbox{Im}(\lambda_{4})h_{S}^{0})
\left[(\chi_{+}^{0})^{3}+3 (\chi_{+}^{0})^{2}\chi_{-}^{0}-3  
\chi_{+}^{0}(\chi_{-}^{0})^{2}
-(\chi_{-}^{0})^{3}\right.\nn\\
& &\left.
+\chi_{+}^{0}(h_{+}^{0})^{2}+\chi_{-}^{0}(h_{+}^{0})^{2}
+2\chi_{+}^{0}h_{+}^{0}h_{-}^{0}-2\chi_{-}^{0}h_{+}^{0}h_{-}^{0}
-\chi_{+}^{0}(h_{-}^{0})^{2}-\chi_{-}^{0}(h_{-}^{0})^{2}\right]\nn\\
& &
+(\mbox{Re}(\lambda_{4})h_{S}^{0}+
\mbox{Im}(\lambda_{4})\chi_{S}^{0})
\left[
(\chi_{+}^{0})^{2}h_{+}^{0}-(\chi_{-}^{0})^{2}h_{+}^{0}
+2 \chi_{+}^{0}\chi_{-}^{0}h_{+}^{0}-
2 \chi_{+}^{0}\chi_{-}^{0}h_{-}^{0}\right.\nn\\
& &\left.
+(h_{+}^{0})^{3}-(h_{-}^{0})^{3}
+(\chi_{+}^{0})^{2}h_{-}^{0}-(\chi_{-}^{0})^{2}h_{-}^{0}
+3(h_{+}^{0})^{2}h_{-}^{0}-3h_{+}^{0}(h_{-}^{0})^{2}\right],\\
\label{v4}
V_{\lambda_{7}H} &=&
\frac{\mbox{Re}(\lambda_{7})}{2}\left[
\{  (\chi_{+}^{0})^{2}+ (\chi_{-}^{0})^{2}  -
(h_{+}^{0})^{2}-(h_{-}^{0})^{2} \}  
\{(\chi_{S}^{0})^{2}-(h_{S}^{0})^{2}\}\right.\nn\\
& &\left.+4\{ \chi_{+}^{0}h_{+}^{0}  +\chi_{-}^{0}h_{-}^{0}   \}
\chi_{S}^{0} h_{S}^{0}\right]
  +\mbox{Im}(\lambda_{7})\left[
(\chi_{+}^{0} h_{+}^{0}+\chi_{-}^{0} h_{-}^{0})
\{(\chi_{S}^{0})^{2}-(h_{S}^{0})^{2}\}\right.\nn\\
& &\left. -\{ (\chi_{+}^{0})^{2}+ (\chi_{-}^{0})^{2}  -
(h_{+}^{0})^{2}-(h_{-}^{0})^{2}   \}
\chi_{S}^{0} h_{S}^{0}\right],
\label{v7}
\ee
where only those terms containing the neutral components are written above.
The rest of the terms in $V_{4H}$ has the form
\be
& &(h_{S}^{0})^{2n_{1}}(h_{+}^{0})^{2n_{2}}(h_{-}^{0})^{2n_{3}}
(\chi_{S}^{0})^{2n_{4}}(\chi_{+}^{0})^{2n_{5}}(\chi_{-}^{0})^{2n_{6}}\\
\label{neutral}
& & \mbox{with}~~\sum_{i=1}^{6}n_{i}=2
~~\mbox{and}~~n_{i}=0,1,2.\nn
\ee

\section{Minimization conditions and Higgs masses}
Below we will analyze the total potential
$V_{T}=V_{H}+V_{SB}$ for the 
three nonequivalent cases (\ref{class1}), (\ref{class2})
and (\ref{class3}). We consider only
phenomenologically viable cases (\ref{require1}).
But we do allow if necessary  a large hierarchy among
the nonvanishing VEVs. In all the cases, $\lambda_4=0$ follows from the discrete symmetry in
question.

\vskip 0.5cm
\noindent
\underline{\bf $R\times S_{2}'$}~
( $\mu_{4}=\mu_{5}=\mu_{6}=0; \lambda_4=0)$:
\vskip 0.5cm
\noindent
The five minimization conditions in this case
are given by
\be
0 &=& -v_{S}\mu_{3}^{2}+\partial V_{4H}/\partial h_{S}^{0},
\label{m31}\\
0 &=& -v_{+}(\mu_{1}^{2}+\mu_{2}^{2})+\partial V_{4H}/\partial h_{+}^{0},
\label{vp1}\\
0 &=&- v_{-}(\mu_{1}^{2}-\mu_{2}^{2})+\partial V_{4H}/\partial h_{-}^{0},
\label{vm1}\\
0 &=& -c_{+}(\mu_{1}^{2}+\mu_{2}^{2})+\partial V_{4H}/\partial \chi_{+}^{0},
\label{cp1}\\
0 &=& -c_{-}(\mu_{1}^{2}-\mu_{2}^{2})+\partial V_{4H}/\partial \chi_{-}^{0},
\label{cm1}
\ee
where the second derivative terms, i.e.,
$\partial V_{4H}/\partial h^{0}~\mbox{and} ~\partial V_{4H}/\partial\chi^{0}$,
are $\sim O(\mbox{VEV}^{3})$.
We first observe that because of the absence of $\lambda_4$
 the condition (\ref{m31}) requires $\mu_{3}\sim O(\mbox{VEV})$.
 If $|\mu_1^2\pm\mu_2^2| >> v^2$ should be satisfied, then
 none of $v_+, v_-, c_+, c_-$ can be $O(v)$.
 But this is not consistent with (\ref{vp1})--(\ref{cm1})
 because of the absence
 of $v_S^3$ terms in the derivative terms of  (\ref{vp1})--(\ref{cm1}).
Therefore, taking into account  the condition (\ref{require1}),
at least one of $v_+, v_-, c_+, c_-$ has to be 
$O(v)$. Assume that $v_{+}\sim O(v)$, which means
that $\mu_{1}^{2}=-\mu_{2}^{2}+O(\mbox{VEV}^{2})$.
Consequently, 
 the total Higgs potential in this case
can be written as
\be
V_{T} &=& -2 \mu_{1}^{2}H_{-}^{\dag}H_{-}+\dots,
\ee
where the terms indicated by $\dots$ are
those which are proportional to  $\mbox{VEV}^{n}~(n=1,\dots,4)$.
Therefore, only $H_{-}$  can obtain
a large mass, if $-2\mu_{1}^{2}$ is positive and large. 
So, this case does not satisfy the phenomenological requirement
that all the physical Higgs bosons except one can be made heavy 
without running into the problem with triviality.

One can perform similar analyses for other cases such as $c_{-}\sim 0(v)$.
[$v_{S} \neq 0$ is always assumed.] As before, one finds that
only one $SU(2)_{L}$ doublet can become heavy.
So, the soft masses  with the discrete symmetry $R\times S_{2}'$
can not be used for a phenomenologically viable  model.

\vskip 0.5cm
\noindent
\underline{\bf $R$}~($\mu_{4}=\mu_{6}=0; \lambda_4=0$):
\vskip 0.5cm
\noindent
The five minimization conditions in this case
are given by
\be
0 &=& -v_{S}\mu_{3}^{2}+\partial V_{4H}/\partial h_{S}^{0}
\label{m32}\\
0 &=&- v_{+}(\mu_{1}^{2}+\mu_{2}^{2})-v_{-}\mu_{5}^{2}
+\partial V_{4H}/\partial h_{+}^{0},
\label{vp2}\\
0 &=&-v_{+}\mu_{5}^{2}-v_{-}(\mu_{1}^{2}-\mu_{2}^{2})
+\partial V_{4H}/\partial h_{-}^{0},
\label{vm2}\\
0 &=&- c_{+}(\mu_{1}^{2}+\mu_{2}^{2})-c_{-}\mu_{5}^{2}
+\partial V_{4H}/\partial \chi_{+}^{0},
\label{cp2}\\
0 &=&-c_{+}\mu_{5}^{2}-c_{-}(\mu_{1}^{2}-\mu_{2}^{2})
+\partial V_{4H}/\partial \chi_{-}^{0}.
\label{cm2}
\ee
Again, because of (\ref{m32}), 
 $\mu_{3}\sim O(\mbox{VEV})$.
 $|\mu_5|$ has to be large, otherwise the situation is the same as
 in the previous case.
Eqs. (\ref{vp2}) and (\ref{vm2}) have a nontrivial solution
\be
\mu_{1}^{2} &=&
-\frac{\mu_{5}^{2}(v_+^2+v_-^2)+O(\mbox{VEV}^{4})}{2v_+v_-},~
\mu_{2}^{2}=\frac{\mu_{5}^{2}(v_+^2-v_-^2)+O(\mbox{VEV}^{4})}{2v_+v_-},
\label{m1m2}
\ee
if $v_+\neq 0, v_- \neq 0$.
 Then the total potential becomes
 \be
 V_{T}&=&  m_{H}^{2}H_{H}^{\dag}H_{H}+\dots,
 \ee
where, as before, 
the terms indicated by $\dots$ are
those which are proportional to  $\mbox{VEV}^{n}~(n=1,\dots,4)$, and 
\be
 H_{H} &=& \frac{v_{-}H_{+}-v_{+}H_{-}}{(v_{+}^{2}+v_{-}^{2})^{1/2}},~
 m_{H} ^{2} =
\frac{v_{+}^{2}+v_{-}^{2}}{v_{+}v_{-}}  \mu_{5}^{2}.
 \ee
 Therefore, only  $H_H$ can become heavy.
 
 If $v_-=0$, Eq.~(\ref{vm2}) requires $|v_+/v| <<1$ because $|\mu_5| >> v$
 has to be satisfied.
 To satisfy Eq.~(\ref{vm2}), on one hand, at least one of $c_+,$ and $c_-$ has to
 be $O(v)$ because of the absence of $v_S^3$ terms in the derivative term.
 On the other hand, 
 we obtain the equation (\ref{m1m2}) with $v_{\pm} \to c_{\pm}$.
 [$c_+\sim O(\mbox{VEV}), c_-=0$ and $c_+=0, c_-= O(\mbox{VEV})$
 can not satisfy (\ref{cp2}) and (\ref{cm2}).]

The case $v_+=0$ is equivalent to the case $v_-=0$.
If $v_+=v_-=0$, the situation does not change.
From these considerations, we conclude that
the  case at hand does not satisfy the phenomenological requirement.

\vskip 0.5cm
\noindent
\underline{\bf $S_{2}'$: ($\mu_{5}=\mu_{6}=0; \lambda_4=0$):}
\vskip 0.5cm
\noindent
The five minimization conditions in this case
are given by
\be
0 &=& -v_{S}\mu_{3}^{2}-\sqrt{2}  v_{+}\mu_{4}^{2}
+\partial V_{4H}/\partial h_{S}^{0},
\label{m33}\\
0 &=& -v_{+}(\mu_{1}^{2}+\mu_{2}^{2})-\sqrt{2}v_{S}\mu_{4}^{2}
+\partial V_{4H}/\partial h_{+}^{0},
\label{vp3}\\
0 &=& -v_{-}(\mu_{1}^{2}-\mu_{2}^{2})
+\partial V_{4H}/\partial h_{-}^{0},
\label{vm3}\\
0 &=& -c_{+}(\mu_{1}^{2}+\mu_{2}^{2})
+\partial V_{4H}/\partial \chi_{+}^{0},
\label{cp3}\\
0 &=&- c_{-}(\mu_{1}^{2}-\mu_{2}^{2})
+\partial V_{4H}/\partial \chi_{-}^{0}.
\label{cm3}
\ee
Note that the derivative terms in  (\ref{vm3})--(\ref{cm3})
contain at least of one of $v_-,c_+$ and $c_-$.
Therefore, large values for $\mu_1$ and $\mu_2$
can be consistent with  (\ref{vm3})--(\ref{cm3}), only if
(i) $v_-=c_+=c_-=0$ and (ii) $\mu_1^2=\mu_2^2+O(\mbox{VEV}^2)$
or (iii) $\mu_1^2=-\mu_2^2+O(\mbox{VEV}^2)$.
Keeping this in mind, 
we next solve (\ref{m33}) and (\ref{vp3}) to obtain
\be
\mu_{3}^{2} &=& 
\frac{v_{+}^{2}(\mu_{1}^{2}+\mu_{2}^{2})+O(\mbox{VEV}^{4})}{v_{S}^{2}},
~\mu_{4}^{2} =
- \frac{v_{+}(\mu_{1}^{2}+\mu_{2}^{2})+O(\mbox{VEV}^{3})}{\sqrt{2}v_{S}}.
\label{mu12}
\ee
Inserting (\ref{mu12}) into the total Higgs  potential $V_{T}$, we obtain
\be
V_{T} &=&
-(\mu_{1}^{2}-\mu_{2}^{2})H_{-}^{\dag}H_{-}
-\frac{\mu_1^2+\mu_2^2}{2v_S^2}\left[
(v_S H_+^{\dag}-v_+ H_S^{\dag})(v_S H_+ -v_+ H_S)+h.c. \right]+\dots
\label{vt2}
\ee
We see from (\ref{vt2}) that the case (ii)  can be ruled out, because in this case
$H_-$ can not obtain a large mass. We can also see from (\ref{vt2}) 
that the case (iii) 
allows large values of the Higgs masses  if $|v_+/v_S| \gsim 40$.
However,  (\ref{vm3}) and (\ref{cm3}) require that $|v_-/v|,|c_-/v| <<1$.
Note that the derivative terms of  (\ref{vm3}) and (\ref{cm3})
contain at least one of $ v_-, c_-$, which implies that
$v_-=c_-=0$ to satisfy (\ref{vm3}) and (\ref{cm3}).
$c_+$ is nonvanishing in the case (iii).
For the case (i) we obtain the same form of the leading potential
$V_T$, but no restriction on the ratio $v_+/v_S$.
In terms of VEVs, we have $v_-=c_+=c_-=0$ for the case (i),
and $v_-=c_-=0$ for (iii). These two types of VEVs 
are $S_2'$ invariant VEVs (\ref{s2vev}).
Both types of VEVs give rise to the general form of the 
ferminon mass matrix (\ref{general-m}).

Below we would like to consider only the case (i) ($v_-=c_+=c_-=0$), and
give  the mass matrix ${\bf m_{h}^2}$  of the  neutral scalar Higgs bosons
\be
h_{-}^{0} &,&
h_L^{0} = \sin\gamma ~h_+^0 +\cos\gamma ~h_S^0,
    h_{H}^{0} =\cos\gamma~ h_{+}^{0}-\sin\gamma ~h_{S}^{0},
    \label{mixh}
   \ee
   and the mass matrix ${\bf m_{\chi}^2}$ the  neutral pseudo scalar Higgs bosons
   \be
 \chi_{-}^{0} &, &    \chi_{L}^{0} =  \sin\gamma ~\chi_{+}^{0}
 +\cos\gamma ~\chi_{S}^{0},
    \chi_{H}^{0} =
  \cos\gamma ~\chi_{+}^{0}-\sin\gamma ~\chi_{S}^{0},
  \label{mixchi}
 \ee
are, respectively, given by
\be
{\bf m_{h}^2} &=&
 \left( \begin{array}{ccc}
m_{h_{-}^{0}}^2 &0 &0 \\
0 & m_{h22}^2 &  m_{h23}^2\\
0 &m_{h23}^2  & m_{h33}^2\\
\end{array}\right),
{\bf m_{\chi}^2} =
 \left( \begin{array}{ccc}
m_{\chi_{-}^{0}}^2 &0 &0 \\
0 & 0 &0 \\
0 & 0 & m_{\chi_{H}^{0}}^2\\
\end{array}\right),
\label{mhiggs}
\ee
where
\be
m_{h_{-}^{0}}^2  &=&2\mu_{2}^{2} +\sqrt{2}\cot\gamma \mu_{4}^{2}
\simeq -(\mu_{1}^{2}-\mu_{2}^{2}),
\label{mhm}\\
m_{h22}^2 & = &v^{2}[~2(\lambda_{1}+\lambda_{3})\sin^{4}\gamma+
\frac{1}{2}(\lambda_{5}+\lambda_{6}+2\lambda_{7})\sin^{2}2\gamma+
2\lambda_{8}\cos^{4}\gamma~]\label{h22},\\
m_{h23}^2 & = &\frac{v^{2}}{2}\sin2\gamma[~(\lambda_{1}+\lambda_{3})(1-\cos  
2\gamma)
+(\lambda_{5}+\lambda_{6}+2\lambda_{7})\cos2\gamma
- 2\lambda_{8}\cos^2\gamma~],\label{h23}\\
m_{h33}^2 & = &2\sqrt{2}\mu_{4}^{2}/\sin2\gamma+
\frac{v^{2}}{2}
(\lambda_{1}+\lambda_{3}-\lambda_{5}-\lambda_{6}
-2\lambda_{7}+\lambda_{8})~\sin^{2}2\gamma,\\
\label{h33}
m_{\chi_{-}^{0}}^2 & = &2\mu_{2}^{2}\sqrt{2}+\mu_{4}^{2}\cot\gamma
-2 (\lambda_{2}+\lambda_{3})v_{+}^{2}-2\lambda_{7}v_{S}^{2},\\
m_{\chi_{H}^{0}}^2 & = &2\sqrt{2}\mu_{4}^{2}/\sin2\gamma-2\lambda_{7}v^{2},
\ee
and we have introduced ($v=(v_{+}^{2}+v_{S}^{2})^{1/2}$)
\be
\tan \gamma &=& \frac{v_+}{v_S}.
\label{gamma}
\ee
In (\ref{mhm})-(\ref{h33}), we have taken into account the higher order 
terms of (\ref{vt2}) 
with $\lambda_{4}=\mbox{Im}(\lambda_{7})=0$.
[$\lambda_4=0$ follows from the $S_2'$ symmetry.]
If $\lambda_{4}$ and $\mbox{Im}(\lambda_{7})$
do not vanish, there is no local minimum for (i).
As we can see from the mass matrices (\ref{mhiggs}) 
with (\ref{mhm})-(\ref{h33}),
the pseudo scalar boson (\ref{mhiggs}), $\chi_{L}$, is the would-be
Goldstone boson, and that except for $h_{L}^{0}$ all the
physical Higgs bosons can become heavy without  large Higgs couplings $\lambda$'s.
We also find from (\ref{mixh}) and (\ref{gamma})  that
only $h_{L}^{0}$ acquires VEV. 
Since only $h_{L}^{0}$ acquires VEV, its coupling to the fermions
is flavor diagonal, while the other physical neutral
Higgs bosons have FCNC couplings.
However, $h_{L}^{0}$ still mixes with
$h_{H}^{0}$ because of the nonvanishing entry $m^{2}_{h23}$.
Therefore, we have to so fine tune that  $m^{2}_{h23}$ vanishes.
(Of course, the mixing is suppressed by $v^{2}/\mu_{4}^{2}
\sim 6\times 10^{-4}$ for $\mu_{4}\sim 10$ TeV.)
In this limit, $m_{h33}$ and $m_{h22}$ 
are the masses of $h_{H}^{0}$ and the lightest Higgs $h_{L}^{0}$,
respectively.

\section{Soft breaking without symmetry}
Here we would like to investigate the full potential
$V_{T}=V_{H}+V_{SB}$ without any assumption on 
abelian discrete symmetries.
The reason is that $S_{2}'$ is not a symmetry of the theory;
it can be  a symmetry only in the Higgs potential.
So, radiative corrections can induce finite non-$S_{2}'$-invariant 
terms in the Higgs potential, for instance.
Here we assume that all the soft masses  (\ref{vSB})
are present and that they are still real.
We, however, do not allow an unnatural large hierarchy of the VEVs, in contrast to
the previous sections.
There are exactly nine nonequivalent possibilities that satisfy the phenomenological
requirement (\ref{require1}):
\be
A_{1} &:& v_{-}=c_{+}=c_{-}=0;~A_{2} : v_{+}=v_{-}=c_{-}=0;\label{1st}\\
B_{1} &:& c_{+}=c_{-}=0;~B_{2} : v_{-}=c_{-}=0;~
B_{3}: v_{-}=c_{+}=0;~B_{4} : v_{+}=v_{-}=0;\label{2nd}\\
C_{1} &:& c_{-}=0;~C_{2}: v_{-}=0;\label{3ird}\\
D &:& \mbox{none of them} ~=0.\label{4th}
\ee
It will turn out that among these nine possibilities
only two cases $A_{1}$ and   $B_{1}$ satisfy the phenomenological
constraint  
that all the Higgs bosons except one can be made heavy 
without running into the problem with triviality.
Note that  $A_{1}$  and also $B_2$ exhibit the $S_{2}'$ 
invariant VEVs (\ref{s2vev}).

\vskip 0.5cm
\noindent
\underline{\bf $A_{1}$}~($ v_{-}=c_{+}=c_{-}=0$):
\vskip 0.5cm
\noindent
We start with the case $A_{1}$.
The first case $A_{1}$ corresponds to the  $S_{2}'$ invariant VEVs (\ref{s2vev}).
The nontrivial minimization conditions at
 $v_{-} =c_{+}=c_{-}=0$ are given by
\be
0&=& 
-v_{S}\mu_{3}^{2}-\sqrt{2}  v_{+}\mu_{4}^{2}+\partial V_{4H}/\partial h_{S}^{0},
\label{m34}\\
0&=& 
-v_{+}(\mu_{1}^{2}+\mu_{2}^{2})-\sqrt{2}v_{S}\mu_{4}^{2}
+\partial V_{4H}/\partial h_{+}^{0},
\label{vp4}\\
0&=& -\sqrt{2}v_{S}\mu_{6}^{2}-v_{+}\mu_{5}^{2}+\partial V_{4H}/\partial h_{-}^{0}
\label{vm4}\\
0&=& 
\partial V_{4H}/\partial \chi_{+}^{0}
=v_{+}v_{S}(v_{+}\mbox{Im}(\lambda_{4})/2\sqrt{2}
+v_{S}\mbox{Im}(\lambda_{7})),
\label{cp4}
\\
0&=& 
\partial V_{4H}/\partial \chi_{-}^{0}=-v_{+}^{2}v_{S}\mbox{Im}(\lambda_{4})/2\sqrt{2}.
\label{cm4}
\ee
(\ref{cp4}) requires $0=\mbox{Im}(\lambda_{7})
+(v_{+}/2\sqrt{2}v_{S})\mbox{Im}(\lambda_{4})$,
and  (\ref{cm4}) requires $\mbox{Im}(\lambda_{4})=0$.
So, we assume that $\lambda_{7}$  and $\lambda_{4}$ are real.
(In the case of the $S_{2}'$ invariant soft term (\ref{class2}), $\lambda_{4}$ has to vanish
for the $S_{2}'$ VEVs (\ref{s2vev}) to correspond to a local minimum.)
We then use (\ref{m34}) --(\ref{vm4}) to express 
$\mu_{1}^{2}, \mu_{3}^{2}$ and  $\mu_{5}^{2}$ in terms of VEVs:
\be
\mu_{1}^{2} &=& -\mu_{2}^{2}-
\sqrt{2}\mu_{4}^{2}\cot\gamma+O(\mbox{VEV}^{2}),~
\mu_{3}^{2} = -\sqrt{2}\mu_{4}^{2}\tan\gamma+O(\mbox{VEV}^{2}),\nn\\
\mu_{5}^{2} &= & -\sqrt{2}\mu_{6}^{2}\cot\gamma+O(\mbox{VEV}^{2}),
\label{mus}
\ee
where $\gamma$ is defined in (\ref{gamma}).
 Inserting $\mu$'s of (\ref{mus}) into the total potential,
 we can compute the mass matrices and find
\be
{\bf m_{h}^2} &\simeq &
 \left( \begin{array}{ccc}
2\mu_{2}^{2}+\sqrt{2}\mu_{4}^{2} \cot\gamma 
 &0 & \sqrt{2}\mu_{6}^{2} /\sin\gamma \\
0 & 0 &  0\\
\sqrt{2}\mu_{6}^{2} /\sin\gamma &0
 &2 \sqrt{2}\mu_{4}^{2} /\sin2\gamma\\
\end{array}\right) +O(\mbox{VEV}^{2})\simeq {\bf m_{\chi}^2}
\label{mhiggs2}
\ee
for the basis (\ref{mixh}) and (\ref{mixchi}).
Comparing these results with (\ref{mhiggs}), we find that apart from the
$O(\mbox{VEV}^{2})$ terms, 
 the masses (\ref{mhiggs2}) reduce to those of the $S_{2}'$ invariant case
 (\ref{mhiggs}) as   $\mu_{6}^{2}$ (and hence  $\mu_{5}^{2}$ because 
 of (\ref{mus})) goes to zero. Therefore, the 
 $S_{2}'$ invariant local minimum exists in the full Higgs potential, if
 all the mass  parameters  are real.
 
\vskip 0.5cm
\noindent
\underline{\bf $A_{2}$}~($ v_{+}=v_{-}=0$):
\vskip 0.5cm
\noindent
 The five minimization conditions at 
 $v_{+} =v_{-}=c_{-}=0$ (which is of the $S_2'$ invariant type (\ref{s2vev})) are given by
\be
0 &=& 
-v_{S}\mu_{3}^{2}+\partial V_{4H}/\partial h_{S}^{0},
\label{m35}\\
0 &=& -\sqrt{2}v_{S}\mu_{4}^{2}
+\partial V_{4H}/\partial h_{+}^{0},
\label{vp5}\\
0 &=&-\sqrt{2}v_{S}\mu_{6}^{2}+\partial V_{4H}/\partial h_{-}^{0}.
\label{vm5}\\
0 &=& -c_{+}(\mu_{1}^{2}+\mu_{2}^{2})+
\partial V_{4H}/\partial \chi_{+}^{0},
\label{cp5}\\
0 & =&-c_{+}\mu_{5}^{2}+
\partial V_{4H}/\partial \chi_{-}^{0}.
\label{cm5}
\ee
 (\ref{m35})- (\ref{cm5}) imply 
 \footnote{As announced, we do not allow an unnatural large hierarchy among the VEVs.
 If, for instance, $|v_S/v| <<1$, then $\mu_3^2$ can be large
 thanks to the nonvanishing $\lambda_4$. 
 In this case, $H_S$ can become heavy.} that 
$\mu_{3}^{2}, (\mu_{1}^{2}+\mu_{2}^{2} ), \mu_{4}^{2},
\mu_{5}^{2} , \mu_{6}^{2}
\sim  O(\mbox{VEV}^{2})$.
Inserting $\mu$'s  above into the total potential, we find
\be
V_{T} &=&2\mu_{2}^{2}H^{\dag}_{-}H_{-}+
 O(\mbox{VEV}^{4}).
 \ee
 So, only $H_{-}$ can become heavy. 
 
 \vskip 0.5cm
\noindent
\underline{\bf $B_{1}$}~($ c_{+}=c_{-}=0$):
\vskip 0.5cm
\noindent
 The five minimization conditions at 
$ c_{+}=c_{-}=0$ are given by
\be
0 &=& 
-v_{S}\mu_{3}^{2}-\sqrt{2}v_{+}\mu_{4}^{2}
-\sqrt{2}v_{-}\mu_{6}^{2}
+\partial V_{4H}/\partial h_{S}^{0},
\label{m36}\\
0 &=& -v_{+}(\mu_{1}^{2}+\mu_{2}^{2})-\sqrt{2}v_{S}\mu_{4}^{2}
-v_{-}\mu_{5}^{2}
+\partial V_{4H}/\partial h_{+}^{0},
\label{vp6}\\
0 &=&-v_{+}\mu_{5}^{2}-\sqrt{2}v_{S}\mu_{6}^{2}
-v_{-}(\mu_{1}^{2}-\mu_{2}^{2})
+\partial V_{4H}/\partial h_{-}^{0},
\label{vm6}\\
0 &=&-(v_{+}^{2} +2 v_{+}v_{-}-v_{-}^{2})v_{S} \mbox{Im} (\lambda_{4})/2\sqrt{2}-
v_{+} v_{S}^{2} \mbox{Im} (\lambda_{7}),
\label{cp6}\\
0 & =&-(v_{+}^{2} -2 v_{+}v_{-}-v_{-}^{2})v_{S} \mbox{Im} (\lambda_{4})/2\sqrt{2}-
v_{-}v_{S}^{2} \mbox{Im}(\lambda_{7}).
\label{cm6}
\ee
(\ref{cp6}) and (\ref{cm6}) require 
$\mbox{Im}(\lambda_{7})=\mbox{Im}(\lambda_{4})=0$.
Solving (\ref{m36}),  (\ref{vp6}) and  (\ref{vm6})
to express $\mu_{1}^{2},\mu_{3}^{2}$ and $\mu_{6}^{2}$
in terms of $v_{S},v_{+}$ and $v_{-}$, and inserting them into the total potential,
we obtain:
\be
V_{T}&=& 
\left[\{2\mu_{2}^{2}v_{-}^{2}+\mu_{5}^{2}
(v_{-}^{3}/v_{+}-v_{+}v_{-})\}/v_{S}^{2})
+\sqrt{2}\mu_{4}^{2}(v_{+}+v_{-}^{2}/v_{+})/v_{S}\right]
H_{S}^{\dag}H_{S}\nn\\
& &+ \left[2\mu_{4}^{2}(v_{S}/v_{+})
+\mu_{5}^{2}(v_{-}/v_{+})\right]H_{+}^{\dag}H_{+}\nn\\
& &+ \left[2\mu_{2}^{2}+\mu_{5}^{2}(v_{-}/v_{+})
+\sqrt{2}\mu_{4}^{2}(v_{S}/v_{+})\right]H_{-}^{\dag}H_{-}\nn\\
& & +\left[-\sqrt{2}\mu_{4}^{2}(v_{-}/v_{+})-2 \mu_{2}^{2}(v_{-}/v_{S})
+\mu_{5}^{2}((v_{+}/v_{S})-(v_{-}^{2}/v_{-}v_{S}))\right]
(H_{S}^{\dag}H_{-}+h.c)\nn\\
& &
-\mu_{5}^{2}(H_{+}^{\dag}H_{-}+h.c)
-\sqrt{2}\mu_{4}^{2}( H_{S}^{\dag}H_{+}+h.c.)
+O(\mbox{VEV}^{4}).
\ee
One can show that except for
$ h_{L} =(v_{S} h_{S}^{0}+v_{+}h_{+}^{0}+v_{-}h_{-}^{0})/(v_{+}^{2}+v_{-}^{2}
+v_{S}^{2})^{1/2}$
all the physical Higgs bosons can become heavy.
So, this case satisfies the phenomenological requirements.

 \vskip 0.5cm
\noindent
\underline{\bf $B_{2,3,4}, C_{1,2}, D$}:
\vskip 0.5cm
\noindent
 We have performed similar analyses for the rest of the cases, and found
 that none of $B_{2,3,4}$, $C_{1,2}$ and $D$ cases satisfy our requirement
 (if we do not allow a large hierarchy among the VEVs).

 \section{The Pakvasa-Sugawara vacuum}
 The Pakvasa-Sugawara (PS) VEVs \cite{pakvasa1} are given by
 \be
 v_{-} &=& c_{+}=0,
 \label{psvev}
 \ee
 which is nothing but the case $B_{3}$ given in (\ref{2nd}).
 As we mentioned, the $S_{3}$ invariant potential (\ref{vH}) does 
 meet the requirement that 
except for one neutral physical Higgs boson, all the physical bosons can become heavy.
On the other hand, the PS VEVs (\ref{psvev}) are the most
economic VEVs in the case of a spontaneous CP violation;
only one phase which should be determined in the Higgs sector
enters into the Yukawa sector.
Here we would like to analyze the most general case with
complex soft masses in contrast to the previous sections.
 The minimization conditions are:
 \be
0 &=& -v_{S}\mu_{3}^{2}-  \sqrt{2}v_{+}\mbox{Re}(\mu_{4}^{2})
+\sqrt{2}c_{-}\mbox{Im}(\mu_{6}^{2})
+\partial V_{4H}/\partial h_{S}^{0},
\label{m37}\\
0 &=& -v_{+}(\mu_{1}^{2}+\mu_{2}^{2})-\sqrt{2}v_{S}\mbox{Re}(\mu_{4}^{2})
+c_{-}\mbox{Im}(\mu_{5}^{2})
+\partial V_{4H}/\partial h_{+}^{0},
\label{vp7}\\
0 &=& -\sqrt{2}v_{S}\mbox{Re}(\mu_{6}^{2})
-v_{+}\mbox{Re}(\mu_{5}^{2})
+\partial V_{4H}/\partial h_{-}^{0},
\label{vm7}\\
0 &=& \sqrt{2}v_{S}\mbox{Im}(\mu_{4}^{2})-c_{-}\mbox{Re}(\mu_{5}^{2})
+\partial V_{4H}/\partial \chi_{+}^{0},
\label{cp7}\\
0 &=&\sqrt{2}v_{S}\mbox{Im}(\mu_{6}^{2})
+v_{+}\mbox{Im}(\mu_{5}^{2})- c_{-}(\mu_{1}^{2}-\mu_{2}^{2})
+\partial V_{4H}/\partial \chi_{-}^{0}.
\label{cm7}
\ee
We solve (\ref{m37})-(\ref{cm7}) to express $\mu_{1}^{2},\mu_{3}^{2}, \mbox{Im}(\mu_{4}^{2}),
\mbox{Re}\mu_{5}^{2}$ and $\mbox{Im}(\mu_{5}^{2})$ in terms of VEVs.
We find that in the leading order, they are given by
\be
\mu_{1}^{2} &=&\left(\mu_{2}^{2}v_{+}^{2}+\sqrt{2}\mbox{Re}(\mu_{4}^{2})v_{+}v_{S}+
\sqrt{2}\mbox{Im}(\mu_{6}^{2})v_{S}c_{-}+\mu_{2}^{2}c_{-}^{2}\right)
/(c_{-}^{2}-v_{+}^{2})\nn\\
& &+O(\mbox{VEV}^{2}),\nn\\
\mu_{3}^{2} &=& \sqrt{2}(-\mbox{Re}(\mu_{4}^{2})(v_{+}/v_{S})
+\mbox{Im}(\mu_{6}^{2})(c_{-}/v_{S}))+O(\mbox{VEV}^{2}),\nn\\
~\mbox{Im}(\mu_{4}^{2})&=-&\mbox{Re}(\mu_{6}^{2})(c_{-}/v_{+})
+O(\mbox{VEV}^{2}),
~\mbox{Re}(\mu_{5}^{2})=-\sqrt{2}\mbox{Re}(\mu_{6}^{2})(v_{S}/v_{+})
+O(\mbox{VEV}^{2}),
\\
\mbox{Im}(\mu_{5}^{2}) &=&\left(\sqrt{2}\mbox{Re}(\mu_{4}^{2})v_{S}c_{-}+
\sqrt{2}\mbox{Im}(\mu_{6}^{2})v_{+}v_{S}+2\mu_{2}^{2}v_{+}c_{-}\right)
/(c_{-}^{2}-v_{+}^{2})\nn\\
& &+O(\mbox{VEV}^{2}).\nn
\ee
Inserting these mass parameters into the full potential, we have verified numerically that 
except for one neutral physical Higgs boson, all the physical bosons can become heavy.
In the limit, in which  the imaginary parts of $\mu_{4}^{2}, \mu_{5}^{2},
\mu_{6}^{2}, \lambda_{4}$ and $\lambda_{7}$ vanish,
the Pakvasa-Sugawara VEVs reduce to the $S_{2}'$ invariant VEVs (\ref{s2vev}), as we can see also  from
\be
c_{-} &\to & \left(-4 \mbox{Im}(\mu_{4}^{2})
+\mbox{Im}(\lambda_{4})v_{+}^{2}+2\sqrt{2}\mbox{Im}(\lambda_{7})v_{+}v_{S}\right)
(v_{+}/4 \mbox{Re}(\mu_{6}^{2}))+\dots,
\ee
where $\dots$ stands for higher orders in the limit.

\section{Supersymmetric extension}
As in the case of the minimal
supersymmetric standard model (MSSM),
we  introduce two $S_3$ doublet Higgs superfields,
$H^U_i, H^D_i (i=1, 2)$, and  two
$S_3$ singlet Higgs superfields, $H^U_S, H^D_S$ \cite{kobayashi1,choi}.
The same R-parity is assigned to these fields as in the MSSM.
Then the most general renormalizable $S_{3}$ invariant
superpotential is given by
\be
W_H &=&
\mu_{1} H^{U}_{i}H^{D}_{i}+\mu_{3} H^{U}_{S}H^{D}_{S}.
\label{poth}
\ee
The $S_{3}$ invariant soft scalar mass terms are  \cite{kobayashi1,choi}:
\be
{\cal L}_{S} &=&
-m_{H_{1}^{U}}^{2} (|\hat{H}_{1}^{U}|^{2}+
|\hat{H}_{2}^{U}|^{2})
-m_{H_{1}^{D}}^{2}(|\hat{H}_{1}^{D}|^{2}
+|\hat{H}_{2}^{D}|^{2})\nn\\
& &
-m_{H_{S}^{U}}^{2}(|\hat{H}_{S}^{U}|^{2})-
m_{H_{S}^{D}}^{2}(|\hat{H}_{S}^{D}|^{2}),
\label{soft3}
\ee
and the $S_{3}$ invariant B terms are:
\be
{\cal L}_{B} &=&B_{1}(\hat{H}_{1}^{U}\hat{H}_{1}^{D}+
\hat{H}_{2}^{U}\hat{H}_{2}^{D})+B_{3}(\hat{H}_{S}^{U}\hat{H}_{S}^{D})
+h.c.,
\label{bterm1}
\ee
where hatted fields are scalar components.
Given the superpotential (\ref{poth}) along with the $S_{3}$ invariant
soft supersymmetry breaking (SSB)  sector 
(\ref{soft3}) and (\ref{bterm1}),
we can now write down the scalar potential.
For simplicity we assume that only the neutral scalar components
of the Higgs supermultiplets acquire VEVs.
The relevant part of the scalar potential is then given by
\be
V &=&
(|\mu_{1}|^{2}+m_{H_{1}^{U}}^{2})(|\hat{H}_{1}^{0U}|^{2}+
|\hat{H}_{2}^{0U}|^{2})+
(|\mu_{1}|^{2}+m_{H_{1}^{D}}^{2})(|\hat{H}_{1}^{0D}|^{2}
+|\hat{H}_{2}^{0D}|^{2})\nn\\
& &+(|\mu_{3}|^{2}+m_{H_{S}^{U}}^{2})(|\hat{H}_{S}^{0U}|^{2})+
(|\mu_{3}|^{2}+m_{H_{S}^{D}}^{2})(|\hat{H}_{S}^{0D}|^{2})\nn\\
& &+\frac{1}{8}(\frac{3}{5}g_{1}^{2}+
g_{2}^{2})(|\hat{H}_{1}^{0U}|^{2}+
|\hat{H}_{2}^{0U}|^{2} +|\hat{H}_{S}^{0U}|^{2}
-|\hat{H}_{1}^{0D}|^{2}-
|\hat{H}_{2}^{0D}|^{2} -|\hat{H}_{S}^{0D}|^{2} )^{2}\nn\\
& &-[~B_{1}(\hat{H}_{1}^{0U}\hat{H}_{1}^{0D}+
\hat{H}_{2}^{0U}\hat{H}_{2}^{0D})+
B_{3}(\hat{H}_{S}^{0U}\hat{H}_{S}^{0D})
+h.c.~],
\label{scalarp1}
\ee
where $g_{1,2}$ are the gauge coupling constants
for the  $U(1)_{Y}$ and $SU(2)_{L}$ gauge groups.
As one can see easily, the scalar potential $V$ (\ref{scalarp1})
has a continues global symmetry $SU(2)\times U(1)$ in addition
to the local $SU(2)_{L}\times U(1)_{Y}$.
As a result, there will be a number of
pseudo Goldstone bosons that are phenomenologically
unacceptable. This is a consequence of $S_{3}$ symmetry.
Therefore, we would like to break $S_{3}$ symmetry explicitly.
As in the non-supersymmetric case,
we would like to break it
as soft as possible to 
preserve predictions from $S_{3}$ symmetry, while
breaking the global  $SU(2)\times U(1)$
symmetry completely.
There is a unique choice for that: Since the softest terms
have the canonical  dimension  two, the soft $S_{3}$ breaking
should be in the SSB sector.
As for the soft scalar masses, we have an 
important consequence (\ref{soft3}) from $S_{3}$ symmetry that
they are diagonal in generations.
Since we would like to preserve this, the only choice is
to introduce  the soft $S_{3}$ breaking 
terms in  the B sector \cite{choi}.
Moreover, looking at the $S_{3}$  invariant scalar potential $V$ (\ref{scalarp1}),
we observe that it has again an abelian discrete symmetry
\be
S_{2}' &:& H_{1}^{U,D} \leftrightarrow H_{2}^{U,D},
\label{s2p2}
\ee
which is the same as (\ref{s2p}).
We assume that the soft $S_{3}$ breaking 
terms respect
this discrete symmetry (\ref{s2p2}), and add the following
 soft $S_{3}$ breaking Lagrangian:
\be
{\cal L}_{S_{3}B} &=&
B_{4}(\hat{H}_{1}^{U}\hat{H}_{2}^{D}
+\hat{H}_{2}^{U}\hat{H}_{1}^{D})
+B_{5}\hat{H}_{S}^{U}(\hat{H}_{1}^{D}+\hat{H}_{2}^{D})
+B_{6}\hat{H}_{S}^{D}(\hat{H}_{1}^{U}+\hat{H}_{2}^{U})+h.c.
\label{s3b}
\ee
In the following discussions, we assume that all the B parameters are real.
The resulting scalar potential can be analyzed, and one finds that
a local minimum respecting 
$S_{2}'$  symmetry, i.e.
\be 
<\hat{H}_{1}^{0U}> &=&<\hat{H}_{2}^{0U}>=v_{U}/2\neq 0~,~
<\hat{H}_{1}^{0D}> =<\hat{H}_{2}^{0D}>=v_{D}/2\neq 0,\nn\\
<\hat{H}_{S}^{0U}> &=& v_{SU}/\sqrt{2}\neq 0~,~
<\hat{H}_{S}^{0D}>= v_{SD}/\sqrt{2}\neq 0,
\label{s2vevs3}
\ee
can occur.
To see this, we write down the minimization conditions in this case,
which can be uniquely solved:
\be
(|\mu_{1}^{2}|^{2} +m^{2}_{H_{1}^{U}})&=& (B_{1}v_{SD}+B_{4} v_{SD}
+\sqrt{2}B_{6} v_{SD})/v_{U}
+O(\mbox{VEV}^{2}),\\
(|\mu_{3}^{2}|^{2} +m^{2}_{H_{S}^{U}}) &=& (B_{3}v_{SD}+\sqrt{2}B_{5} v_{D})
/v_{SU}+O(\mbox{VEV}^{2}),\\
(|\mu_{1}^{2}|^{2} +m^{2}_{H_{1}^{D}}) &=& (B_{1}v_{U}+B_{4} v_{U}
+\sqrt{2}B_{5} v_{SU})/v_{D}+O(\mbox{VEV}^{2}),\\
(|\mu_{3}^{2}|^{2} +m^{2}_{H_{S}^{D}}) &= & (B_{3}v_{SU}+\sqrt{2}B_{6} v_{U})/v_{SD}
+O(\mbox{VEV}^{2}),
\ee
Inserting these solutions
into the scalar potential (\ref{scalarp1}) with (\ref{s3b}), 
we obtain the mass matrices
for the Higgs fields. As in the non-supersymmetric case, we redefine
the Higgs fields as
\be
\hat{H}_{\pm}^{D,U} &=&\frac{1}{\sqrt{2}}(
\hat{H}_{1}^{D,U}\pm\hat{H}_{2}^{D,U}).
\ee
Then the mass matrices can be written as
\be
{\bf M}^{2}_{-} = \left( \begin{array}{cc}
((B_{1}+B_{4})v_{D}+\sqrt{2} B_{6}v_{SD})/v_{U} &-B_{1}+B_{4}
\\  -B_{1}+B_{4} & ((B_{1}+B_{4})v_{U}+\sqrt{2} B_{5}v_{SU})/v_{D}
\end{array}\right)+O(\mbox{VEV}^{2})
\label{Mm}
\ee
for the $(~\hat{H}^{U}_{-} , (\hat{H}^{D}_{-})^{\dag}~)$ basis, and 
\be
{\bf M}^{2} = \left( \begin{array}{cccc}
M_{US} &0 & -B_{3} &-\sqrt{2}B_{5}
\\ 
0 & M_{U+} & -\sqrt{2}B_{6}  & -B_{1}-B_{4}
\\
-B_{3} & -\sqrt{2}B_{6} & M_{DS} & 0
\\
-\sqrt{2}B_{5}&- B_{1}-B_{4} & 0  & M_{D+}
\end{array}\right)+O(\mbox{VEV}^{2})
\label{Mp}
\ee
for the 
$(~\hat{H}^{U}_{S} , \hat{H}^{U}_{+},
(\hat{H}^{D}_{S})^{\dag},(\hat{H}^{D}_{+})^{\dag}~)$ basis,
where
\be
M_{US} &=&( B_{3}v_{SD}+\sqrt{2}B_{5}v_{D})/v_{SU},~
M_{U+}=(( B_{1}+B_{4})v_{D}+\sqrt{2}B_{6}v_{SD})/v_{U},\\
M_{DS} &=&( B_{3}v_{SU}+\sqrt{2}B_{6}v_{U})/v_{SD},~
M_{D+} =(( B_{1}+B_{4})v_{U}+\sqrt{2}B_{5}v_{SU})/v_{D}.
\ee
From the mass matrices (\ref{Mm}) and (\ref{Mp}), we find
that
the lightest physical Higgs boson, 
the MSSM Higgs boson, can be written as a linear combination
\be
h_{\rm MSSM} &=&
(v_{D}\hat{H}_{+}^{0D}+v_{SD}\hat{H}_{S}^{0D}+
v_{U}\hat{H}_{+}^{0U}+v_{SU}\hat{H}_{S}^{0U})/v,
\ee
where
$v=(v_{U}^{2}+v_{SU}^{2}+v_{D}^{2}+v_{SD}^{2})^{1/2}
\simeq 246 ~\mbox{GeV}$, 
and its mass is approximately given by
\be
m_{h} &\simeq &((3/5)g_{1}^{2}+g_{2}^{2})
(v_{U}^{2}+v_{SU}^{2}-v_{D}^{2}-v_{SD}^{2})^{2}/v
\label{mh}
\ee
for $\mu^{2\prime}s , B's >> v^{2}$. It can be shown that the masses 
of  the other physical Higgs bosons can be made
arbitrary heavy. 
From (\ref{mh}) we see that the tree-level upper bound
for $m_{h}$ is exactly the same as in the MSSM.

Because of the very nature of the SSB terms, the explicit breaking of $S_{3}$
in the B sector (\ref{s3b})
does not propagate to the other sector.
Moreover, although the superpotential (\ref{poth})
and the corresponding trilinear couplings 
do not respect $S_{2}'$ symmetry (\ref{s2p2}),
they can not generate $S_{2}'$
violating infinite B terms  because 
they  can generate only $S_{3}$ invariant
terms, which are however automatically $S_{2}'$ invariant.

\section{Conclusions}
We  recall that our investigations have been carried out under
the two phenomenological conditions (\ref{constraint1}) and
(\ref{require1}).
Below we would like to summarize our conclusions.\\

\vspace*{0.5cm}
\noindent
1: The $S_{3}$ invariant Higgs potential (\ref{vH}) does not satisfy the
phenomenological\\ requirement that

\vspace{0.3cm}
\noindent
``except one neutral physical Higgs boson all the physical Higgs bosons can
become heavy $\gsim 10$ TeV without having a problem
with triviality. ''

\vspace{0.3cm}
\noindent
That is, for a phenomenological viable model
we have to break $S_{3}$ explicitly  if we do not introduce further Higgs fields.

\vspace*{0.5cm}
\noindent
2: Among the  real nonequivalent soft $S_{3}$ breaking
 masses (\ref{class1}), (\ref{class2}) and (\ref{class3}) that
can be characterized according to  discrete symmetries,
only the $S_{2}'$ invariant case (\ref{class2}) with 
the $S_{2}'$ invariant VEVs (\ref{s2vev})  can satisfy the phenomenological
requirement of 1. 

\vspace*{0.5cm}
\noindent
3:
Even for  the most general
quartic Higgs potential 
 with the most general real $S_{3}$ breaking
 masses (\ref{vSB}),
the $S_{2}'$ invariant VEVs (\ref{s2vev})  can 
correspond to  a local minimum and 
 satisfy the phenomenological
requirement of 1.

\vspace*{0.5cm}
\noindent
4:
The Pakvasa-Sugawara VEVs (\ref{psvev}) can be a local
minimum in the case of the most general
quartic Higgs potential 
 with the most general complex $S_{3}$ breaking masses and  can
satisfy the phenomenological
requirement of 1.

\vspace*{0.5cm}
\noindent
5: In  a minimal supersymmetric extension 
with the $S_{2}'$ invariant, real soft $S_{3}$
breaking masses in the B sector,
the phenomenological
requirement of 1 can be satisfied with the  $S_{2}'$ invariant
VEVs (\ref{s2vevs3}),
where the other B parameters are also assumed to be real.
These B terms violate supersymmetry as well as $S_{3}$ softly.
This possibility to introduce 
$S_{2}$ violating soft terms in the B sector only 
 is consistent with renormalizability.
The lower bound of the lightest Higgs boson is the same as in the MSSM.

\vspace*{0.5cm}
It is a very difficult task to test the Higgs  sector experimentally.
However, as we see from (\ref{vt2}) and (\ref{mhiggs}), in the case of the $S_{2}'$
 invariant soft breaking with the $S_{2}'$ invariant VEVs (\ref{s2vev}),
there are basically only two masses
$m_{h_{H}}$ and $m_{h_{-}}$ for four neutral and 
two  charged heavy Higgs bosons. This may be experimentally tested,
because their couplings to the fermions are fixed \cite{kubo1,kubo2}.

\acknowledgments
 One of us (JK) would like to thank 
 Manuel Drees and Manfred Lindner for useful discussions, and Walter Grimus for 
useful discussions and his hospitality at the Universit\" at Wien.
This work is supported by the Grants-in-Aid for Scientific Research 
from the Japan Society for the Promotion of Science
(No. 13135210).



\end{document}